\documentclass[journal]{IEEEtran}
\IEEEoverridecommandlockouts
\usepackage{type1ec}
\usepackage[T1]{fontenc}
\usepackage{cite}
\usepackage{amsmath,amssymb}

\usepackage{algorithmic}
\usepackage{float} 
\usepackage{pstricks}
\usepackage{pstricks-add}
\usepackage{pst-circ}
\usepackage{subcaption}
\usepackage{graphicx}
\usepackage{textcomp}
\usepackage{xcolor}
\usepackage{pst-coil}
\usepackage{pst-node}
\usepackage{pst-all}	
\usepackage{pst-func}
\usepackage{array}
\usepackage{placeins}
\usepackage{siunitx} 
\def\Tframe#1{\TR{\psframebox{#1}}}
\def\BibTeX{{\rm B\kern-.05em{\sc i\kern-.025em b}\kern-.08em
    T\kern-.1667em\lower.7ex\hbox{E}\kern-.125emX}}
\usepackage[ruled,vlined]{algorithm2e}
\usepackage{slashbox}
\usepackage{mathtools}
\usepackage{multirow}
\usepackage[bookmarks=false]{hyperref}
\usepackage{url}
\usepackage{verbatim}

\begin{document}
\begin{NoHyper}

\title{Tb/s Polar Successive Cancellation Decoder \\ 16nm  ASIC Implementation
}
\author{Altu\u{g}~S\"{u}ral, E. G\"{o}ksu~Sezer, Ertu\u{g}rul  Kola\u{g}as{\i}o\u{g}lu, Veerle Derudder and Kaoutar Bertrand
\thanks{A. S\"{u}ral, E. G. Sezer and E. Kola\u{g}as{\i}o\u{g}lu are with POLARAN, Ankara, Turkey (e-mails: \{altug.sural, goksu.sezer, ekolagasioglu\}@polaran.com).}
\thanks{V. Derudder and K. Bertrand are with IMEC, Leuven, Belgium (e-mails: \{veerle.derudder, kaoutar.bertrand\}@imec.be).}
\thanks{Manuscript received September 15, 2020}} 
\maketitle
\begin{abstract} 
This work presents an efficient ASIC implementation of successive cancellation (SC) decoder for polar codes. SC is a low-complexity depth-first search decoding algorithm, favorable for beyond-5G applications that require extremely high throughput and low power. The ASIC implementation of SC in this work exploits many techniques including pipelining and unrolling to achieve Tb/s data throughput without compromising power and area metrics. To reduce the complexity of the implementation, an adaptive log-likelihood ratio (LLR) quantization scheme is used. This scheme optimizes bit precision of the internal LLRs within the range of 1-5 bits by considering irregular polarization and entropy of LLR distribution in SC decoder. The performance cost of this scheme is less than 0.2 dB when the code block length is 1024 bits and the payload is 854 bits. Furthermore, some computations in SC take large space with high degree of parallelization while others take longer time steps. To optimize these computations and reduce both memory and latency, register reduction/balancing (R-RB) method is used. The final decoder architecture is called optimized polar SC (OPSC). The post-placement-routing results at 16nm FinFet ASIC technology show that OPSC decoder achieves 1.2 Tb/s coded throughput on 0.79 mm$^2$ area with 0.95 pJ/bit energy efficiency.

\end{abstract}

\IEEEpeerreviewmaketitle
\section{Introduction} \label{sec:intro}
The Ethernet Alliance foresees a strong demand for Tb/s data throughput for data centers and mobile networks \cite{ethernetroadmap2019}. In the wireless domain, the Horizon 2020 project \textit{Enabling Practical Wireless Tb/s Communications with Next Generation Channel Coding} (EPIC) \cite{epic} considers three well-known forward error correction (FEC) schemes turbo \cite{Berrou1993}, Low Density Parity Check (LDPC) \cite{Gallager1962}, \cite{MacKay1996} and polar codes \cite{Arikan2009} for extremely high speed beyond 5G applications. This paper aims to present an efficient polar code implementation to meet Tb/s throughput demand. Polar codes have attracted a significant interest from both academia and industry and recently they have been adopted for the protection of the control channel in the enhanced mobile broadband (eMBB) service for the fifth generation (5G) cellular wireless technology \cite{gppNR}. 

Polar codes are a unique family of FEC schemes which can theoretically achieve capacity in broad class of channels using a low-complexity successive cancellation (SC) decoder \cite{Arikan2009}, \cite{Sasaoglu2009}. To improve error correction performance of SC algorithm at moderate data block lengths, many algorithms are proposed with most popular one being SC-list (SCL) \cite{Tal2015}. SCL algorithm can track a list of possible decision candidates and can achieve near ML performance with this additional complexity. Other SC based decoding algorithms SC-flip (SCF) \cite{Afisiadis2014} and Soft-cancellation (SCAN) \cite{Fayyaz2014} use an iterative approach to correct decision errors of SC. Using multiple iterations or multiple parallel decoders make these algorithms much more power hungry at Tb/s data rates.

The sequential processing nature of SC limits parallelism within decoder but promotes pipelined approach for Tb/s throughput. There are some high-throughput and pipelined SC implementations \cite{Zhang2013}, \cite{Dizdar2016}, \cite{Giard2015}, \cite{Giard2016} in literature. These implementations can only achieve a few Gb/s throughput, even if the unrolled implementations \cite{Giard2015}, \cite{Giard2016} take advantage of a set of shortcuts \cite{Yazdi2011}, \cite{Sarkis2014}, \cite{Hanif2017} for speeding up the SC decoder. For a higher throughput, discrete quantization of soft-information \cite{Shah2019} and register reduction/balancing (R-RB) \cite{Sural2019} methods have been proposed. In \cite{Sural2019}, Tb/s throughput for polar SC decoder have been investigated. A generic problem identified for Tb/s throughput regime is power density caused by excessive switching activity in a limited core area. 


This work proposes an optimized SC (OPSC) decoder architecture based on pipelining and unrolling techniques. The OPSC decoder has low implementation complexity thanks to careful register balancing and adaptive log-likelihood ratio (LLR) quantization (AQ) scheme. This scheme takes LLR distribution of each polar code segment as input and optimizes bit precision of internal LLRs. In addition to that OPSC utilizes R-RB method to optimize clock frequency by flattening time delay of pipeline stages. The post-placement-routing (post-P\&R) results at 16nm FinFet ASIC technology show that OPSC decoder achieves 1.2 Tb/s coded throughput (corresponds to 1 Tb/s data throughput) on 0.79 mm$^2$ area with 0.95 pJ/bit energy efficiency.

\subsection{Summary of the achievements} \label{sec:achievements}
\begin{itemize}
\item OPSC decoder and a channel simulator capable of simulating very low error rates are implemented on FPGA to verify RTL code and measure error correction performance at very low BER. The FPGA implementation results show that OPSC decoder achieves 200 Gb/s throughput and $1.1 \times 10^{-13}$ bit error rate (BER) at 8 dB Eb/No.
\item OPSC decoder exploits AQ and R-RB methods to reduce design area and power.
\item To the best of our knowledge, OPSC decoder is the first polar decoder that exceeds 1 Tb/s throughput on ASIC based on post-P\&R results. 
\item The results also show that OPSC decoder archives 0.95 pJ/bit energy efficiency at 0.79 mm$^2$ area.
\end{itemize}

The outline of this paper is as follows. Section \ref{sec:polar_codes_and_sc} gives a short summary of polar coding and introduces the SC decoding with shortcuts and AQ. Section \ref{sec:architecture_general} presents the proposed decoder architecture including pipeline depth optimization. Section \ref{sec:fpga_verification} presents FPGA verification and communication performance of the proposed decoder. Section \ref{sec:asic_implementation} presents ASIC implementation details and show comparison with state-of-the-art implementations. Finally, Section \ref{sec:conclusion} summarizes the main results with a brief conclusion.

\section{Polar Codes and SC Decoding Algorithm}  \label{sec:polar_codes_and_sc}
\subsection{Polar codes}   \label{sec:polar_code_review}
Polar codes are a class of linear block codes that exists with a block length $N=2^n$ for every $n\ge 1$. A polar transform matrix $G_N = G^{\otimes n}$ is the $n^{\text{th}}$ Kronecker power of a generator matrix $G =\begin{bmatrix}
1 & 0 \\
1 & 1 
\end{bmatrix}$. An input transform vector $u_N$ consists of a user data set $u_{\cal A}$ to carry the user data $d_K$ with $K$ dimention and a redundancy (frozen) set $u_{{\cal A}^c}$ to carry the frozen bits fixed to zero. The polar-encoded codeword $x_N$ is simply obtained as $x_N = u_N G_N$. We refer to \cite{Arikan2009} for a complete description of the polar coding technique. 

\subsection{SC Decoding Algorithm with Shortcuts}  \label{sec:sc_algorithm}
The data flow diagram of SC algorithm with certain shortcuts is shown in Fig. \ref{fig:sc_shortcut_diagram}. At the start of decoding, channel log-likelihood ratio (LLR) vector $\ell_N$ is given to the input. For an AWGN channel $W$ with the variance $\sigma^2$,  i$^{\text{th}}$ ($1  \leq i \leq N$) encoded symbol $x_i$ and received symbol $y_i$, a channel LLR vector $\ell_i$ is

\begin{align}
\ell_i &= \ln \left( \frac{W(y_i|x_i=0)}{W(y_i|x_i=1)} \right) \nonumber \\
	&= \ln \left( \frac{e^{\frac{-(y_i-1)^2}{2\sigma^2}}}{\sqrt{2\pi\sigma^2}}\right) - \ln \left( \frac{e^{\frac{-(y_i+1)^2}{2\sigma^2}}}{\sqrt{2\pi\sigma^2}} \right) \nonumber \\  
	&= \frac{-(y_i-1)^2}{2\sigma^2}-\frac{-(y_i+1)^2}{2\sigma^2} \nonumber \\
	&= \frac{2y_i}{\sigma^2}\text{,} \nonumber 
\end{align}
where $W(y_i|x_i)$ is the channel transition probability density function. 
The forward LLR calculation module consists of F and G functions to calculate the inputs $\ell_M$ for the first and the second half of a polar code, where $M$ is the recursive block length parameter. Initially, $M=N$ and there are $M/2$ size-$2$ $\text{F}_2$ and $\text{G}_2$ functions (denoted as $\text{F}_{M/2}$ and $\text{G}_{M/2}$) at each recursion.
The function $\text{F}_2(\ell_1, \ell_2)$ for any two LLR values $\ell_1$ and $\ell_2$ is defined as 
\begin{align}
 	\text{F}_2(\ell_1, \ell_2) = \operatorname{sgn}(\ell_1  \ell_2)\operatorname{min}(|\ell_1|,|\ell_2|) \text{.} \nonumber
\end{align}
The function $\text{G}_2(\ell_1, \ell_2, \hat{z})$ with a hard decision (HD) feedback $\hat{z}$ is 
\begin{align}
	\text{G}_2(\ell_1, \ell_2, \hat{z}) &= {(1-2\hat{z})}{\ell_1} + \ell_2  \text{.} \nonumber
\end{align}
After a bunch of F and G function iterations, the SC decoder becomes ready for making hard decisions using certain shortcuts. These shortcuts are named as Rate-0 (R0), Rate-1 (R1), SPC and REP (first introduced in \cite{Sarkis2014}) for easily decodable polar code segments. For R0 shortcut all values are assigned to 0. For R1 shortcut a simple threshold function is used to assign 0 for positive LLRs and 1 for negative LLRs. Moreover, Wagner \cite{Silverman1954} and MAP \cite{Fossorier1999} decoders are employed for SPC and REP nodes, respectively. After one of these shortcuts is activated, the backward HD module (can also be named as partial-sum update logic -  PSUL) takes the HD estimate $\hat{u}_M$ and calculates the feedback $\hat{z}_{M/2}$ for the G functions. This module utilizes $M/2$ XOR ($\oplus$) functions at each iteration. After all polar code segments are decoded, the final HD estimate $\hat{u}_N$ is calculated. and the estimated user data $\hat{d}_K$ is extracted from $\hat{u}_N$ at the end of decoding operation.

\begin{figure} 
\centering
\begin{pspicture}(0,0)(8,7)
\rput(0,-1){
\rput(5.7,1.8){$\hat{d}_K$}
\rput(1,1.8){$\hat{u}_N$}
\rput(2.3,7.5){$\ell_N$}
\rput(1,6.2){$\hat{z}_{M/2}$}
\rput(7,7){\fontsize{7pt}{7pt}$M=M/2$}
\rput(7,7.4){$\ell_M$}
\rput(4,6.25){$\ell_M$}
\rput(5.2,6.2){{\fontsize{7pt}{7pt}$M=M/2$}} 
\rput(2.6,3.75){$\hat{u}_M$}
\rput(1.3,3.7){{\fontsize{7pt}{7pt}$M=2M$}} 
\rput(3.45,4.7){\fontsize{8pt}{8pt}$\hat{u}_M$}

\psline[linestyle=dashed,linewidth=0.03](4,4.2)(3.8,3.5)
\psline[linestyle=dashed,linewidth=0.03](4.4,4.2)(4.6,3.5)
\rput(1,0){
\psline[linestyle=dashed,linewidth=0.03](4,4.2)(3.8,3.5)
\psline[linestyle=dashed,linewidth=0.03](4.4,4.2)(4.6,3.5)}
\rput(2,0){
\psline[linestyle=dashed,linewidth=0.03](4,4.2)(3.8,3.5)
\psline[linestyle=dashed,linewidth=0.03](4.4,4.2)(4.6,3.5)}
\rput(3,0){
\psline[linestyle=dashed,linewidth=0.03](4,4.2)(3.8,3.5)
\psline[linestyle=dashed,linewidth=0.03](4.4,4.2)(4.6,3.5)}

\rput(0.2,-1.4){
\psframe(3.6,3.7)(4.4,4.9)
\psline{->}(4,3.8)(4,4.8)\psline{<->}(3.6,3.8)(4.4,3.8)
\psline[linestyle=dashed,linewidth=0.02](3.6,4)(4.4,4)
\rput(4.2,4.2){\fontsize{6pt}{6pt}$0$}}
\rput(-1.8,-1.4){
\psframe(6.6,3.7)(7.4,4.9)
\psline{->}(7,3.8)(7,4.8)\psline{<->}(6.6,3.8)(7.4,3.8)
\psline[linestyle=dashed,linewidth=0.02](7,4.4)(6.6,4.4)\psline[linestyle=dashed,linewidth=0.02](7,4)(7.4,4)
\rput(7.2,4.4){\fontsize{6pt}{6pt}$1$}\rput(6.8,4){\fontsize{6pt}{6pt}$0$}}
\psframe(5.8,3.5)(6.6,2.3) \rput(6.2,2.9){\begin{tabular}{c}
\fontsize{6pt}{6pt}$\text{Wagner}$ \\ {\fontsize{6pt}{6pt}$\text{Decoder}$}\end{tabular}}
\rput(1,0){
\psframe(5.8,3.5)(6.6,2.3) \rput(6.2,2.9){\begin{tabular}{c}
\fontsize{6pt}{6pt}$\text{MAP}$ \\ {\fontsize{6pt}{6pt}$\text{Decoder}$}\end{tabular}}}

\rput(0,0.3){
\psellipse(0.8,7)(0.8,0.4)\rput(0.8,7){Input}
\psline{->}(1.6,7)(3.5,7)}

\rput(6.4,-1.5){
\psellipse(0.8,3)(0.8,0.4)\rput(0.8,3){Output}}

\rput(0,-1){
\psframe(2,2)(5,3) \rput(3.5,2.5){User data extraction}}

\psline(6.1,7.2)(8,7.2)
\psline(8,7.215)(8,4.5)
\psline{->}(8.015,4.5)(7.7,4.5)
\psline{->}(3.7,4.5)(3.2,4.5)
\psline(0.6,4.8)(0.1,4.8)
\psline(0.1,4.785)(0.1,6)
\psline(0.085,6)(2,6)
\psline(2,5.985)(2,7)
\psline{->}(1.985,7)(3.5,7)

\psline(0.6,4.5)(0.1,4.5)
\psline(0.1,4.515)(0.1,1.5)
\psline{->}(0.085,1.5)(2,1.5)
\psline{->}(5,1.5)(6.4,1.5)

\rput(-2.2,5.75){
\rput(7,2){Forward LLR calc.}
\rput(6,1.05){G} \rput(8,1.05){G}
\psframe(5.8,0.85)(6.2,1.25) \psframe(7.8,0.85)(8.2,1.25)
\rput(6,1.5){F} \rput(8,1.5){F} 
\psframe(5.8,1.3)(6.2,1.7) \psframe(7.8,1.3)(8.2,1.7)
\rput(7,1.25){\fontsize{9pt}{9pt}$M/2$}
\psline[linestyle=dotted,linewidth=0.08](6.3,1.05)(7.7,1.05)
\psline[linestyle=dotted,linewidth=0.08](6.3,1.5)(7.7,1.5)
\psframe(5.7,0.75)(8.3,1.8) 
\psline(8.3,1)(8.5,1)
\psline(8.5,1.015)(8.5,0.3)
\psline(8.515,0.3)(5.5,0.3)
\psline(5.5,0.285)(5.5,1.015)
\psline{->}(5.5,1)(5.7,1)
\rput(7,0.1){Iterate until shortcut}}

\rput(3.7,4){
\rput(2,1.2){Hard-decision (HD) making}
\psframe(0,0)(4,1) 
\psframe(0.1,0.2)(0.9,0.8) \rput(0.5,0.5){R0}
\rput(1,0){\psframe(0.1,0.2)(0.9,0.8) \rput(0.5,0.5){R1}}
\rput(2,0){\psframe(0.1,0.2)(0.9,0.8) \rput(0.5,0.5){SPC}}
\rput(3,0){\psframe(0.1,0.2)(0.9,0.8) \rput(0.5,0.5){REP}}}

\rput(-5.1,3.25){
\rput(7,2){Backward HD calc.}
\rput(2,-3){\pscircle(4,4){0.2} \psline(3.9,4)(4.1,4) \psline(4,3.9)(4,4.1)} 
\rput(4,-3){\pscircle(4,4){0.2} \psline(3.9,4)(4.1,4) \psline(4,3.9)(4,4.1)} 
\rput(2,-2.5){\pscircle(4,4){0.2} \psline(3.9,4)(4.1,4) \psline(4,3.9)(4,4.1)} 
\rput(4,-2.5){\pscircle(4,4){0.2} \psline(3.9,4)(4.1,4) \psline(4,3.9)(4,4.1)} 
\rput(7,1.25){\fontsize{9pt}{9pt}$M/2$}
\psline[linestyle=dotted,linewidth=0.08](6.3,1.01)(7.7,1.01)
\psline[linestyle=dotted,linewidth=0.08](6.3,1.5)(7.7,1.5)
\psframe(5.7,0.75)(8.3,1.8) 
\psline{<-}(8.3,1)(8.5,1)
\psline(8.5,1.015)(8.5,0.3)
\psline(8.515,0.3)(5.5,0.3)
\psline(5.5,0.285)(5.5,1.015)
\psline(5.5,1)(5.7,1)
\rput(7,0.1){Iterate until feedback}}
}

\end{pspicture}
\caption{Data flow diagram of polar SC decoding algorithm with shortcuts}
\label{fig:sc_shortcut_diagram}
\end{figure}
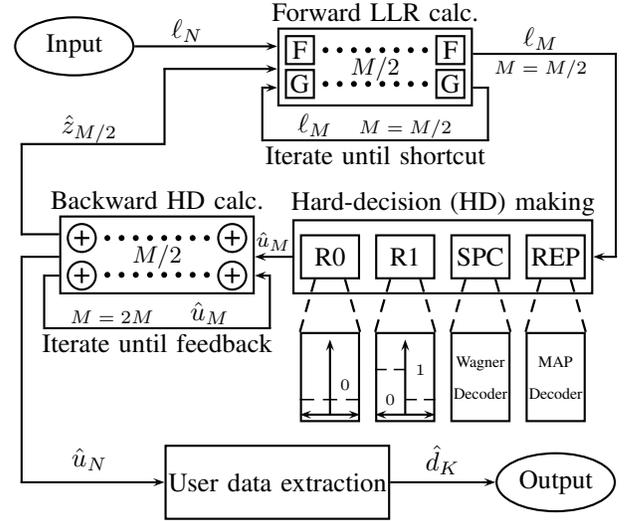

A formal representation of SC algorithm with shortcuts is shown in Algorithm \ref{sc_algorithm}. $v_M$ is the indicator vector of the frozen coordinates for length-$M$ polar code segments. The i$^\text{th}$ element of $v_M$ is defined as
\begin{equation} 
 v_i  = 
	\begin{cases} 1, & \mbox{if } i \in {\cal A}^c \nonumber \\ 
	0, & \mbox{if }i \in {\cal A} \text{.} \nonumber
	\end{cases} 
\end{equation}
When $v_M$ is all one, the R0 shortcut is calculated. When $v_M$ is all zero, the R1 shortcut is calculated using a threshold function on LLRs. 
Both F and G functions are used element-wise for odd $\ell_M^{\text{odd}}$ and even $\ell_M^{\text{even}}$ elements of $\ell_M$ vector. 

\SetKwData{Left}{left}\SetKwData{This}{this}\SetKwData{Up}{up}
\SetKwFunction{Union}{Union}\SetKwFunction{FindCompress}{FindCompress}
\SetKwInOut{Input}{input}\SetKwInOut{Output}{output}
\SetArgSty{textnormal}
\begin{algorithm}
\renewcommand{\algorithmicrequire}{\textbf{Inputs   :}}
\renewcommand{\algorithmicensure}{\textbf{Output:}}
\algorithmicrequire{ $\ell_M$, $v_M$, $M$} \text{  } \algorithmicensure{ $\hat{u}_M$}\\
  \uIf(\tcp*[f]{R0, R=0}){$v_M= 1$ }{
  $\hat{u}_M  = \text{d}(\ell_M$, $v_M= 1) = 0$\\
  }
  \uElseIf(\tcp*[f]{R1, R=1}){$v_M=0$ }{
  $\hat{u}_M =\text{d}(\ell_M$, $v_M= 0)$  \\
  }
  \uElseIf{$M \leq N_{\text{LIM}}$ \bf{and} $v_1= 1$  \bf{and} $v_2^M= 0$  }{
  $\hat{u}_M = \text{d}(\ell_M$, $v_M= 0)$  \tcp*[f]{Wagner dec.} \\ 
  $p = \text{mod}(\sum_{i=1}^M\hat{u}_i,2)$ \tcp*[f]{R = (M-1)/M} \\
  $r = \text{argmin}(\mid \ell_M \mid$)\\
  $\hat{u}_r = \hat{u}_r \oplus p$\\
  }
  \uElseIf{$M \leq N_{\text{LIM}}$ \bf{and} $v_1^{M-1}= 1$  \bf{and} $v_M=0$ }{
  $\hat{u}_M = \text{d}(\sum_{i=1}^M\ell_i, v = 0)$  \tcp*[f]{MAP dec. R = 1/M}  \\
  }
  \Else(\tcp*[f]{Conventional SC $\forall$ R}){ 
   $l_{M/2} = \text{F}(\ell_M^{\text{odd}}$, $\ell_M^{\text{even}}$)  \tcp*[f]{Fig. \ref{fig:f2}} \\
   $\hat{z}_{M/2} = \text{SC}(l_{M/2}$, $v_M^{\text{odd}}$, $\frac{M}{2})$\\
   $r_{M/2} = \text{G}(\ell_M^{\text{odd}}$, $\ell_M^{\text{even}}$, $\hat{z}_{M/2})$ \tcp*[f]{Fig. \ref{fig:g2}} \\
   $\hat{x}_{M/2} = \text{SC}(r_{M/2}$, $v_M^{\text{even}}$, $\frac{M}{2}$)\\
   $\hat{u}_M^{\text{odd}} = \hat{z}_{M/2}$ $\oplus$ $\hat{x}_{M/2}$ \tcp*[f]{Backward HD}\\
   $\hat{u}_M^{\text{even}} = \hat{x}_{M/2}$\\
   }
   \uIf(\tcp*[f]{User data extraction}){$M = N$ }{
	 $\hat{u}_K = u_{\cal A}$ \\
	 \Return {$\hat{u}_K$}
   }
   \Else(\tcp*[f]{Decode remaining code segments}){ 
     \Return {$\hat{u}_M$}
   }
\caption{SC with shortcuts}
\label{sc_algorithm}
\end{algorithm}

\subsubsection{Shortcuts for ($N=1024,K=854$) polar code} \label{sec:shortcuts}
For a specific implementation of polar codes, code parameters are selected as $N=1024$, $K=854$, $R= \frac{5}{6}$ . The selected polar code is constructed using Density Evolution algorithm \cite{Tal2013} at 6.5 dB target Es/No. The number of shortcuts for this code is shown in Table \ref{table:list_of_shortcuts}. Due to high code rate, R1 and SPC shortcuts appear more frequent than R0 and REP shortcuts. SPC and REP shortcuts are not allowed to be greater than $N_{\text{LIM}} = 32$ by design choice to keep the target clock frequency high. There are also other shortcuts discovered in \cite{Hanif2017}, however such shortcuts are not employed in this work due to negligible hardware gain in our target polar code.

\begin{table}[ht]
\centering
\caption{Number of shortcuts in (1024,854) OPSC decoder }
\label{table:list_of_shortcuts}
\begin{tabular}{cccccc}
\hline
\multirow{2}{*}{Node Length} & \multicolumn{5}{c}{Shortcuts} \\ 
 & R0 & R1 & SPC & REP & Total \\ \hline
2   & 2 & 2 & - & - & 4 \\ 
4   & 2 & 4 & 8 & 10 & 24 \\ 
8   & 1 & 3 & 6 & 3 & 13 \\ 
16  & 1 & 3 & 3 & 2 & 9 \\ 
32  & 1 & 6 & 4 & - & 3 \\ 
64  & - & 3 & - & - & 7 \\ 
128 & - & 1 & - & - & 1 \\ 
Total & 68 & 604 & 256 & 96 & 1024 \\ \hline
\end{tabular}
\end{table}


Due to using shortcuts, F/G function and XOR gate gains are shown in Table \ref{table:fg_gain}. In standard SC decoder architecture, there are $N/2=512$ F$_2$ and G$_2$ functions at each polar code segment. Therefore, total number of required F$_2$ and G$_2$ functions are $N \log N= 10,240$. It reduces to 5276 after applying shortcuts. The gain is mostly caused by the smaller polar code segments. The number of F$_2$ functions are not equal to G$_2$ functions for small segments due to R0 shortcuts. For these specific nodes, G functions are used with all zero decision feedback. Furthermore, the required XOR gates for PSUL reduces from $N/2 \log N= 5120$ to 2672.

\begin{table}[ht]
\centering
\caption{F/G function and XOR gate gain due to shortcuts}
\label{table:fg_gain}
\begin{tabular}{ccccc}
\hline
\multirow{2}{*}{\begin{tabular}[c]{@{}c@{}}Node \\ Length\end{tabular}} & \multicolumn{3}{c}{Number of F/G functions} & \multirow{2}{*}{\begin{tabular}[c]{@{}c@{}}XOR gates\\ in PSUL\end{tabular}} \\ 
 & F & G & Total &  \\ \hline
4 & - & 2 & 4 & 2 \\ 
8 & 11 & 13 & 96 & 13 \\ 
16 & 12 & 13 & 200 & 13 \\ 
32 & 10 & 11 & 336 & 11 \\ 
64 & 10 & 11 & 672 & 11 \\ 
128 & 7 & 7 & 896 & 7 \\ 
256 & 4 & 4 & 1024 & 4 \\ 
512 & 2 & 2 & 1024 & 2 \\ 
1024 & 1 & 1 & 1024 & 1 \\ 
Total & 2604 & 2672 & 5276 & 2672 \\ \hline
\end{tabular}
\end{table}

\subsubsection{Adaptive quantization of the LLRs} \label{sec:adapt_quantization}
Adaptive quantization is an optimization method to reduce hardware complexity by decreasing LLR bit precision of internal data paths in the SC decoder. Instead of storing and processing LLRs with a constant number of bits, a variable number of bits is used for each polar code segment. During SC decoding of polar codes, the polarization phenomenon becomes effective. As polarization increases, resolution can be decreased without losing performance and representing polarized code segments with constant quantization bits becomes inefficient. As polarization increases reliability, resolution of LLR can be decreased. Unlike using lookup tables as in \cite{Shah2019} for the computation of LLRs, we use input LLR distribution statistics of each polar code segment and apply the given F and G functions with an optimized bit precision. Adaptive quantization scheme for (1024,854) polar code is shown in Fig. \ref{fig:adaptive_quantization}. The number of quantization bits are written on the lines between polar code segments. With the adaptive quantization, the memory complexity is reduced by 15.1\% while having 4.25 bit average LLR bit precision compared to the SC decoder with 5-bit regular quantization levels. It further reduces combinational logic complexity and enables shallower pipeline depth.

\begin{figure}[htbp]
\centering
\psscalebox{0.65}{
\psset{arrowscale=1}
\psset{unit=0.8cm}
\psset{xunit=1,yunit=1}
\begin{pspicture}(-2,-1.1)(12,11.1)
\rput(5,5){
\psset{arrows=->}
\pstree[treemode=R,levelsep=20ex]{\Tframe{(1024,854)}}
       {\pstree[treemode=R]{\Tframe{(512,361)}\taput{5}}
               {\pstree[treemode=R]{\Tframe{(256,131)}\taput{5}}
                       {\Tframe{(128,36) }\taput{5}
                        \Tframe{(128,95) }\taput{4}
                       }
                \pstree[treemode=R]{\Tframe{(256,230)}\taput{4}}
                       {\Tframe{(128,103)}\taput{4}
                        \Tframe{(128,127)}\taput{3}
                       }
               }
        \pstree[treemode=R]{\Tframe{(512,493)}\trput{4}}
               {\pstree[treemode=R]{\Tframe{(256,238)}\taput{4}}
                       {\Tframe{(128,111)}\taput{4}
                        \Tframe{(128,127)}\taput{3}
                       }
                \pstree[treemode=R]{\Tframe{(256,255)}\taput{3}}
                       {\Tframe{(128,127)}\taput{3}
                        \Tframe{(128,128)}\taput{1}
                       }
               }
        }
		}
\end{pspicture}
}
	\caption{Adaptive quantization of the constituent codes of SC(1024,854) for $128 \leq M \leq 1024$. The number of quantization bits are written on the lines.}
	\label{fig:adaptive_quantization}
\end{figure}
\section{OPSC Decoder Architecture} \label{sec:architecture_general}
The proposed OPSC decoder architecture exploits pipelining and unrolling techniques to achieve Tb/s data rate while keeping implementation complexity in check.
The enabling methods for OPSC decoder are as follows.
\begin{itemize}
\item Systematic polar code for improved BER performance.
\item Min-sum decoding for simpler arithmetic operations with small area and power dissipation.
\item Adaptive quantization of internal LLRs to reduce computational and memory complexity.
\item Bit-reversed order computation to operate on neighboring LLRs.
\item Unrolled and pipelined architecture for high throughput.
\item Fully-parallel SC architecture with multi-bit hard decisions using shortcuts.
\item R-RB using pipeline depth optimization for minimum delay and power. Pipeline depth of OPSC decoder is optimized as 158 for FPGA and 60 for ASIC.
\end{itemize}

The OPSC decoder denoted as $\text{OPSC}(N,K)$, consists of two sub-decoders which have the same block length $\frac{N}{2}$ with a different payload $K_i = \frac{N}{2} R_i$. In general, $\text{OPSC}(N,K)$ is decoded in four steps: calculation of F functions, $\text{OPSC}_1(\frac{N}{2},K_1)$, calculation of G functions and $\text{OPSC}_2(\frac{N}{2},K_2)$. AQ is applied after F and G  functions to reduce LLR bit resolution from Q to Q' bits. For example, the recursive OPSC decoder architecture for $N=16$, $K=9$ polar code segment is shown in Fig. \ref{fig:sc_architecture}. Choice of lower layer code rates is exemplary. The $\ell_{16}$ LLRs at the input with $16 \times Q$ bits are stored during processing duration of $\text{OPSC}(8,2)$ decoder (denoted as $\mathcal{L}(\text{OPSC}_1)$) until $\hat{z}_8$ becomes ready at the input of function block G. Likewise, $\hat{z}_8$ is stored until $\hat{x}_8$ is ready. Buffer memory structure is used to access data faster than other alternatives such as SRAM. 

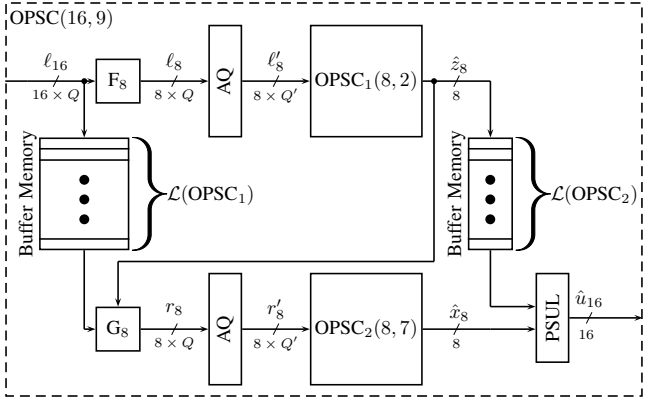
\begin{figure}[htbp]
\centering
	\psscalebox{0.75}{
	\begin{pspicture}(-0.4,1.4)(10.3,8.4)
	\rput(-0.3,0){
	\psdots[linecolor=black,dotsize=3pt](1,7)
	\psdots[linecolor=black,dotsize=3pt](7.2,7)
	\psline{->}(0,7)(1.2,7)
	\rput(0.6,8.1){$\text{OPSC}(16,9)$}
	\rput(0.5,7.35){$\ell_{16}$}
	\rput(0.5,6.75){\fontsize{7pt}{7pt}\selectfont$16\times{Q}$} 
	\psline[linewidth=0.5pt](0.45,6.9)(0.55,7.1)
	\psframe(1.2,6.6)(2,7.4)
	\rput(1.6,7){F$_8$}
	\psline{->}(1,7)(1,6)
	\psframe(0.2,6)(1.8,4)
	\psline(0.2,5.8)(1.8,5.8)
	\psline(0.2,5.6)(1.8,5.6)
	\psline(0.2,4.2)(1.8,4.2)
	\psdots[linecolor=black,dotsize=5pt](1,4.55)
	\psdots[linecolor=black,dotsize=5pt](1,4.9)
	\psdots[linecolor=black,dotsize=5pt](1,5.25)
	\rput{90}(0,5){Buffer Memory}
	\psline(1,4)(1,2.6)
	\psline{->}(1,2.6)(1.2,2.6)
	\psframe(1.2,2.2)(2,3)
	\rput(1.6,2.6){G$_8$}
	\psline{->}(2,7)(3.2,7)
	\psline[linewidth=0.5pt](2.55,6.9)(2.65,7.1)
	\rput(2.6,7.35){$\ell_{8}$}
	\rput(2.6,6.75){\fontsize{7pt}{7pt}\selectfont$8\times{Q}$} 
	\psframe(3.2,6)(3.8,8)
	\rput{90}(3.5,7){AQ}
	\psline{->}(2,2.6)(3.2,2.6)
	\psline[linewidth=0.5pt](2.55,2.5)(2.65,2.7)
	\rput(2.6,2.95){$r_{8}$}
	\rput(2.6,2.35){\fontsize{7pt}{7pt}\selectfont$8\times{Q}$} 
	\rput(7.6,2.35){\fontsize{7pt}{7pt}\selectfont$8$}
	\rput(7.6,6.75){\fontsize{7pt}{7pt}\selectfont$8$}
	\rput(9.9,2.5){\fontsize{7pt}{7pt}\selectfont$16$}
	\psframe(3.2,1.6)(3.8,3.6)
	\rput{90}(3.5,2.6){AQ}
	\psline{->}(3.8,7)(5,7)
	\rput(4.4,7.35){$\ell'_{8}$}
	\rput(4.4,6.75){\fontsize{7pt}{7pt}\selectfont$8\times{Q'}$} 
	\psline[linewidth=0.5pt](4.35,6.9)(4.45,7.1)
	\psline[linewidth=0.5pt](4.35,2.5)(4.45,2.7)
	\rput(4.4,2.95){$r'_{8}$}
	\rput(4.4,2.35){\fontsize{7pt}{7pt}\selectfont$8\times{Q'}$} 
	\psline{->}(3.8,2.6)(5,2.6)
	\psframe(5,6)(7,8)
	\rput(6,7){$\text{OPSC}_1(8,2)$}
	\psframe(5,1.6)(7,3.6)
	\rput(6,2.6){$\text{OPSC}_2(8,7)$}
	\psline(7,7)(8.2,7)
	\psline(7.2,7)(7.2,3.8)
	\psline(7.2,3.8)(1.6,3.8)
	\psline{->}(1.6,3.8)(1.6,3)
	\rput(7.65,7.3){$\hat{z}_8$}
	\psline[linewidth=0.5pt](7.55,6.9)(7.65,7.1)
	\psline{->}(8.2,7)(8.2,6)
	\psline(7.8,5.8)(8.6,5.8)
	\psline(7.8,5.6)(8.6,5.6)
	\psline(7.8,4.2)(8.6,4.2)
	\psframe(7.8,4)(8.6,6)
	\psdots[linecolor=black,dotsize=5pt](8.2,4.55)
	\psdots[linecolor=black,dotsize=5pt](8.2,4.9)
	\psdots[linecolor=black,dotsize=5pt](8.2,5.25)
	\rput{90}(7.6,5){Buffer Memory} 
	\psline(7,2.6)(8.2,2.6)
	\psline(8.2,4)(8.2,3)
	\psline{->}(8.2,2.6)(9,2.6)
	\psline{->}(8.2,3)(9,3)  
	\rput(7.65,2.9){$\hat{x}_8$}
	\psline[linewidth=0.5pt](7.55,2.5)(7.65,2.7)   
	\psframe(9,2)(9.6,3.6)
	\rput{90}(9.3,2.8){PSUL}
	\psline[linewidth=0.5pt](9.9,2.7)(10,2.9) 
	\rput(9.95,3.15){$\hat{u}_{16}$} 
	\psbrace[nodesepB=5.5pt](1.85,4)(1.85,6){$\mathcal{L}(\text{OPSC}_{1})$}
	\psbrace[nodesepB=5.5pt](8.65,4)(8.65,6){$\mathcal{L}(\text{OPSC}_{2})$}      
	\psframe[linestyle=dashed](-0.4,1.4)(10.9,8.4) 
	\psline(0,7)(-0.4,7)
	\psline{->}(9.6,2.8)(10.9,2.8) }  
	\end{pspicture}
	}
    \caption{OPSC(16,9) architecture}
	\label{fig:sc_architecture}
\end{figure}

Similar hardware architecture is utilized to decode all polar code segments recursively. When shortcuts are detected, $\text{OPSC}_1$ and $\text{OPSC}_2$ blocks are replaced. For example, $\text{OPSC}_2(8,7)$ is replaced with an SPC shortcut.

\subsection{Building blocks of the OPSC decoder architecture} \label{sec:architecture}

The basic building blocks of OPSC decoder are F and G functions. F$_N$ function consists of $N/2$ copies of F$_2$ function shown in Fig. \ref{fig:f2}. The F$_2$ function contains a compare-and-select (C\&S) logic and an XOR gate.
\begin{figure}[ht]
\centering
\begin{pspicture}(0.5,2.5)(5,5)
\psset{unit=0.5cm}
\rput(0,1){
\rput(0,2){\logicxor[ninputs=2]{0}(5,5){}}
\psset{unit=1cm}
\psline{->}(0.8,2.8)(3,2.8)
\psline{->}(0.5,2.2)(3,2.2)
\psline{->}(0,4.25)(3,4.25)
\psline{->}(0,3.75)(3,3.75)
\psline(4.25,4)(5.6,4)
\psline(0.8,4.25)(0.8,2.8)
\psline(0.5,3.75)(0.5,2.2)
\rput(0.3,4.45){$\ell_1$}
\rput(0.3,3.95){$\ell_2$}
\psdots[linecolor=black,dotsize=3pt](0.8,4.25)
\psdots[linecolor=black,dotsize=3pt](0.5,3.75)	
\rput(1.8,4.45){$\text{sgn}(\ell_1)$}
\rput(1.8,3.95){$\text{sgn}(\ell_2)$}
\rput(4.92,4.2){$\text{sgn}(\ell_3)$}
\rput(2,3){$\left|\ell_1\right|$}
\rput(2,2.4){$\left|\ell_2\right|$}
\rput(5,2.75){$\left|\ell_3\right|$}
\psline(5.6,2.5)(5.6,4)
\psline(4,2.5)(5.6,2.5)
\psframe(3,2)(4,3)\rput(3.5,2.5){C$\&$S}
\psdots[linecolor=black,dotsize=3pt](5.6,4)
\psline{->}(5.6,4)(6.3,4)
\rput(6,4.2){$\ell_3$}
}
\end{pspicture}
\caption{F$_2$ function.}\label{fig:f2}
\end{figure}
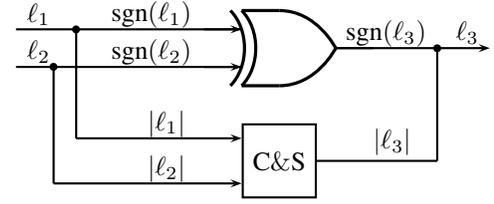

G$_2$ function is shown in Fig. \ref{fig:g2}. The G$_2$ function contains an adder, a subtractor and a multiplexer. The select input $\hat{z}$ of the multiplexer may have longer delay than the $\ell_1' + \ell_2'$  and $\ell_2' - \ell_1'$ inputs due to XOR gate chain in PSUL. To avoid timing problems, both results are calculated and the correct one is chosen. Since LLRs are stored in sign-magnitude form, the G$_2$ function also utilizes two sign-magnitude to twos-complement converters (S2C) at the input and a twos-complement to sign-magnitude converter (C2S) at the output.
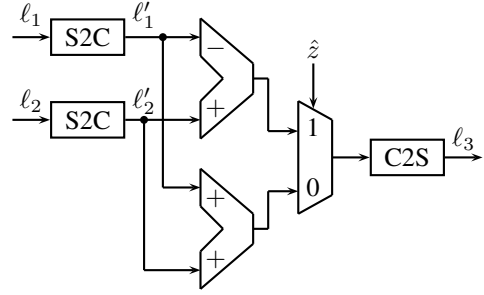
\begin{figure}[ht]
\centering
\begin{pspicture}(0.5,4.5)(6,8)
\rput(0,3.5){
\psline{->}(0,4.35)(0.5,4.35)
\psline{->}(0,3.25)(0.5,3.25)
\rput(0.25,4.65){$\ell_1$}
\rput(0.25,3.5){$\ell_2$}
\rput(-0.5,1.5){\psframe(1,1.5)(2,2)\rput(1.5,1.75){S2C}}
\rput(-0.5,2.6){\psframe(1,1.5)(2,2)\rput(1.5,1.75){S2C}}
\rput(1.75,4.65){$\ell_1'$}
\rput(1.75,3.5){$\ell_2'$}
\rmultiput{
\psline(1.5,1.5)(1.5,1)
\psline(1.5,2.6)(2.2,2)
\psline(1.5,2.1)(1.5,2.6)
\psline(1.5,2.1)(1.8,1.8)
\psline(1.5,1.5)(1.8,1.8)
\psline(1.5,1)(2.2,1.6)
\psline(2.2,1.6)(2.2,2)}(1,0)(1,2)
\psline{->}(1.5,4.35)(2.5,4.35)
\psline{->}(1.5,3.25)(2.5,3.25)
\psline(1.75,3.25)(1.75,1.25)
\psline(2,4.35)(2,2.35)
\psline{->}(1.75,1.25)(2.5,1.25)
\psline{->}(2,2.35)(2.5,2.35)
\psdots[linecolor=black,dotsize=3pt](2,4.35)
\psdots[linecolor=black,dotsize=3pt](1.75,3.25)
\psline(3.2,1.8)(3.4,1.8)\psline(3.4,3.8)(3.4,3.1)\psline{->}(3.4,3.1)(3.8,3.1)
\psline(3.2,3.8)(3.4,3.8)\psline(3.4,1.8)(3.4,2.3)\psline{->}(3.4,2.3)(3.8,2.3)
\rput(0,-0.25){
\rput(4,3.4){1}
\rput(4,2.6){0}
\psline(3.8,2.25)(3.8,3.75)
\psline(3.8,2.25)(4.25,2.5)
\psline(3.8,3.75)(4.25,3.5)
\psline(4.25,2.5)(4.25,3.5)}
\psline{->}(4,4)(4,3.4)\rput(4,4.2){$\hat{z}$}
\psline{->}(4.25,2.75)(4.75,2.75)
\psframe(4.75,2.5)(5.75,3)\rput(5.25,2.75){C2S}
\psline{->}(5.75,2.75)(6.25,2.75)\rput(6,3){$\ell_3$}
\rput(2.7,4.2){$-$}
\rput(2.7,3.4){$+$}
\rput(2.7,2.2){$+$}
\rput(2.7,1.4){$+$}
}
\end{pspicture}
\caption{G$_2$ function.}\label{fig:g2}
\end{figure}
\subsubsection{Complexity analysis} \label{sec:memory_complexity}
The time complexity of fully-parallel standard SC decoder is
\begin{align}
T_\text{N} &= 2 T_{\text{N}/2} + 2 \nonumber = \sum_{i=1}^n 2^{i} = 2N-2 =\Theta (N) \text{.}\nonumber 
\end{align}

The memory complexity of unrolled and fully-pipelined standard SC decoder is
\begin{align}
M_\text{N} &= 2 M_{\text{N}/2} + (Q+0.5)(N^2-N)  +  NQ \nonumber \\
           &=(Q+0.5)\big((2-2^{-\log N}) N^2 - N \log N - N\big)  \nonumber \\
           &\qquad + QN\log N  \nonumber \\
           &= \Theta (N^2 Q) \nonumber \text{.}
\end{align}
where $M_2 = 2Q + 1$. This formula shows that in the most general case, memory complexity increases almost quadratically with $N$. This memory is dominantly used in buffers, where soft decision values are stored. Size of these buffers decreases significantly after applying AQ and R-RB as shown in Table \ref{table:memory_details}. The final memory complexity of OPSC decoder is significantly smaller than the conventional SC decoder.
\subsection{Register reduction/balancing} \label{sec:register_balancing}
Pipeline stages are important to shorten the critical path and, thus, increase the clock frequency. However, excessive pipeline stages may also increase memory complexity. To reduce the pipeline depth, we merge a set of consecutive short paths of the SC decoder as much as possible based on the combinational delay from timing simulations. Register reduction is challenging for SC decoding algorithm due to its sequential essence. We overcome this problem by estimating delay of each computation and exploiting shortcuts for parallel processing. This method enables the decoder to perform multiple calculations within a single clock cycle with remarkably reduced latency and memory consumption. Table \ref{table:memory_details} shows that LLR buffer memory of OPSC decoder is reduced from 1.1 Mb to 380 Kb using R-RB at $Q=5$ bits. After applying R-RB, the pipeline depth becomes 60. The results include shortcuts without AQ. Applying the proposed AQ scheme in Fig. \ref{fig:adaptive_quantization} further reduces the LLR buffer memory to 361 Kb. Including PSUL memory, total buffer memory becomes 380 Kb. The memory gain of AQ is marginal, because R-RB scheme has already reduced the memory significantly.

\begin{table}[ht]
\centering
\caption{Buffer memory of OPSC decoder}
\label{table:memory_details}
\begin{tabular}{ccccc}
\hline
\multirow{2}{*}{\begin{tabular}[c]{@{}c@{}}Node \\ Length\end{tabular}} & \multicolumn{2}{c}{\begin{tabular}[c]{@{}c@{}}Buffer memory\\ depth w/o R-RB\end{tabular}} & \multicolumn{2}{c}{\begin{tabular}[c]{@{}c@{}}Buffer memory\\ depth with R-RB \end{tabular}} \\ 
 & LLR & PSUL & LLR & PSUL \\ \hline
4 & - & 18 & - & - \\ 
8 & 22 & 29 & - & 5 \\ 
16 & 44 & 36 & 15 & 14 \\ 
32 & 60 & 44 & 20 & 15 \\ 
64 & 76 & 50 & 29 & 22 \\ 
128 & 93 & 49 & 29 & 21 \\ 
256 & 103 & 53 & 34 & 20 \\ 
512 & 103 & 46 & 37 & 17 \\ 
1024 & 112 & - & 41 & - \\ 
Total size & 1.1 Mb & 49 Kb & 380 Kb & 19 Kb \\ \hline
\end{tabular}
\end{table}

\section{FPGA Verification and Performance} \label{sec:fpga_verification}
Error correction performance of OPSC decoder was verified on Xilinx (xcvu37p-fsch2892-2L-e-es1) FPGA demo board. To attain real-time verification, FPGA architecture is developed for both polar systematic encoder and OPSC decoder. The rest of this section presents FPGA test platform and error correction performance of OPSC decoder.
\subsection{FPGA test platform} \label{sec:fpga_test_platform}
Polar decoder implementations were verified on FPGA test platform shown in Fig. \ref{fig:fpga_test}. The test platform supports 200 Gb/s information throughput at 234 MHz clock frequency. A linear feedback shift register (LFSR) array generates $K=854$ bit pseudo random data for each transmitted polar codeword. A systematic polar encoder generates $N=1024$ bit encoded data from the pseudo random data. The encoded data, $x_i$, consists of 854 systematic bits and 170 parity bits, where both bits are mapped to BPSK symbols ($s_i$) using the mapping rule in Eq. \ref{bpsk_map_rule}. The symbols are accumulated with the additive white Gaussian noise (AWGN) generated by a build-in Gaussian random number generator. BPSK demodulator generates LLR values with $Q=5$ bits in the form of sign-magnitude. The polar SC decoder processes LLRs and produces information bit estimates, which are compared with the original data to produce error statistics. 


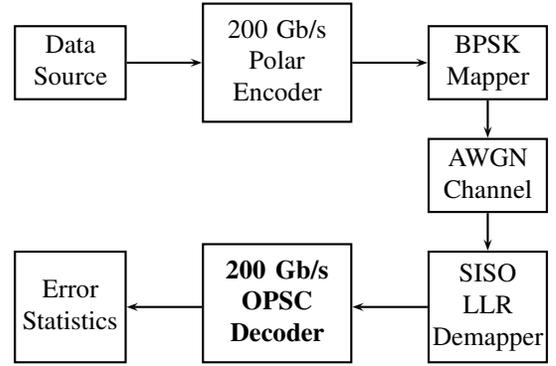
\begin{figure} 
\centering
\begin{pspicture}(0,0)(7.5,5)
\psframe(0,3.5)(1.5,4.5)\rput(0.75,4){\begin{tabular}{c}
Data\\Source \end{tabular}}
\psline{->}(1.5,4)(2.5,4)
\psframe(2.5,3.2)(4.5,4.8)\rput(3.5,4){\begin{tabular}{c}
200 Gb/s \\Polar \\ Encoder \end{tabular}}
\psline{->}(4.5,4)(5.5,4)
\psframe(5.5,3.5)(7.1,4.5)\rput(6.3,4){\begin{tabular}{c}
BPSK\\Mapper \end{tabular}}
\psline{->}(6.3,3.5)(6.3,3)
\psframe(5.5,2)(7.1,3)\rput(6.3,2.5){\begin{tabular}{c}
AWGN\\Channel \end{tabular}}
\psline{->}(6.3,2)(6.3,1.5)
\psframe(5.5,0)(7.1,1.5)\rput(6.3,0.75){\begin{tabular}{c}
SISO \\LLR \\ Demapper \end{tabular}}
\psline{->}(5.5,0.75)(4.5,0.75)
\psframe(2.5,0)(4.5,1.6)\rput(3.5,0.8){\begin{tabular}{c}
\textbf{200 Gb/s} \\ \textbf{OPSC} \\ \textbf{Decoder} \end{tabular}}
\psline{->}(2.5,0.75)(1.5,0.75)
\psframe(0,0)(1.5,1.5)\rput(0.75,0.75){\begin{tabular}{c}
Error\\Statistics \end{tabular}}
\end{pspicture}
\caption{Block diagram of FPGA test platform}\label{fig:fpga_test}
\end{figure}

\begin{equation} 
\label{bpsk_map_rule}
 s_i  = 
	\begin{cases} 
	1, & \mbox{if } x_i = 0     \\ 
	-1, & \mbox{otherwise.}
	\end{cases} 
\end{equation}
\subsection{FPGA performance results} \label{sec:fpga_performance}
The frame error statistics in Fig. \ref{fig:sc_fer} show that the FPGA implementation of OPSC decoder causes less than 0.1 dB performance loss compared to floating-point software simulation (without AQ). The performance loss is caused by 5-bit quantization of LLRs on FPGA. The proposed AQ scheme causes almost 0.1 dB more performance loss, which is tolerable due to hardware implementation gains. 
\begin{figure}[ht]
	\centering
	\includegraphics[scale=0.7]{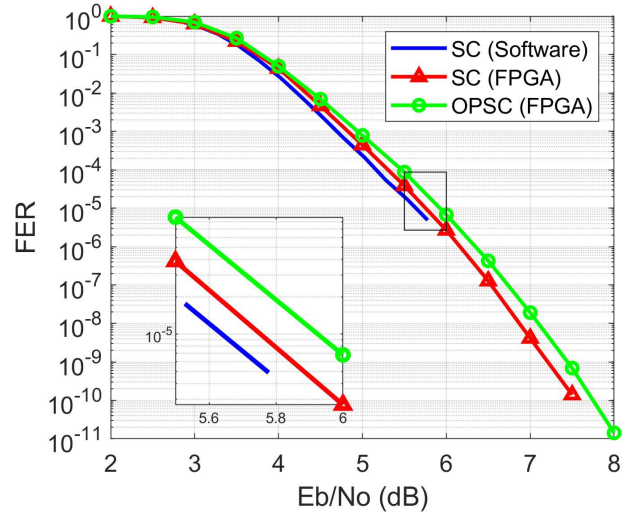}
	\caption{FER performance of (1024,854) polar code}
	\label{fig:sc_fer}
\end{figure}

BER performance results in Fig. \ref{fig:sc_ber} show that a coding gain of $13.93 - 7.72 = 6.21$ dB is attained at $10^{-12}$ BER relative to uncoded transmission.
\begin{figure}[ht]
	\centering
	\includegraphics[scale=0.7]{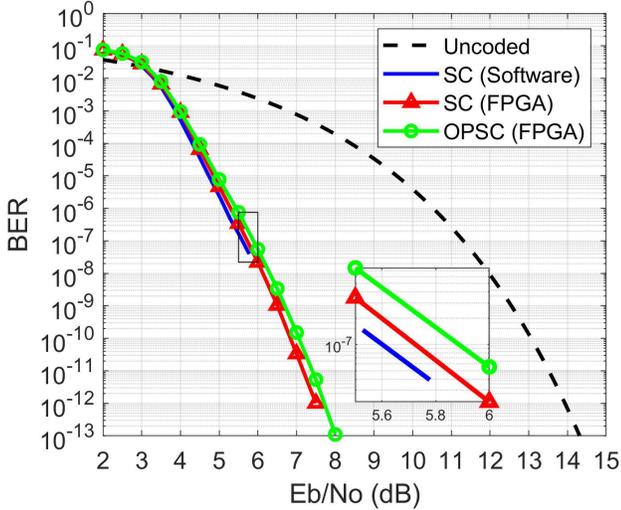}
	\caption{BER performance of (1024,854) polar code}
	\label{fig:sc_ber}
\end{figure}

FPGA implementation results of polar encoder and OPSC decoder are shown in Table \ref{table:resource_utilization_fpga}. Both encoder and decoder have 234 MHz clock frequency. To support this frequency, OPSC decoder utilizes register array memory in the form of LUT for storing internal LLRs. The received LLRs are stored in 143 block random access memory (BRAM) with 16K capacity. The results also show that OPSC decoder utilizes only 7.34 \% of the FPGA in terms of LUT consumption. OPSC decoder can fit low-cost FPGA boards such as Xilinx Artix-7. 

\begin{table}[ht]
\centering
\caption{FPGA resource utilization}
\label{table:resource_utilization_fpga}
\begin{tabular}{ccccc}
\hline
 & LUT & FF & Power (mW) & Latency (ns) \\ \cline{2-5}
Polar encoder & 1848 & 1825 & 36 & 8.5 \\ 
OPSC decoder & 95653 & 50843 & 2373 & 672 \\ \hline
\end{tabular}
\end{table}

\FloatBarrier
\section{ASIC Implementation} \label{sec:asic_implementation}

This section presents ASIC implementation procedure and results. The general information about the ASIC implementation is as follows.
\begin{itemize}
\item The TSMC 16nm CMOS logic FinFet library (BWP16P90) is used for RTL synthesis and backend implementation.
\item The process-voltage-temperature (PVT) values are 0.8 V and \ang{25}C. Although timing is satisfied at 0.7 V at the end of RTL synthesis, 0.8 V is chosen for backend implementation to make timing closure easier.
\item The logic cells with a set of thresholds RVT, LVT, OPT-LVT and ULVT are used.
\item The setup and hold time of each design are verified for typical, worst-C, worst-RC, best-C and best-RC design corners.
\item The power results are estimated accurately by generating signal activity factors using a testbench. The testbench simulates consecutive decoding of 1000 codewords transmitted at 0-9dB Eb/No. For each Eb/No value 100 codewords are tested. 
\item To achieve timing-clean results, a noticeable number of buffers and inverters have been added to the design. 
\item The final implementation results are obtained at the end of a timing-clean P\&R.
\end{itemize}

\subsection{Synthesis} \label{sec:synthesis}

The 0.8V TSMC 16nm logic FinFET plus 1P13M process is used for implementation. Physical synthesis is performed with Cadence Genus. The OPSC has been implemented for a clock frequency of 1.2 GHz. To cope with the high clock frequency, retiming has been adopted to move the pipeline registers across the combinatorial logic. In addition, during synthesis fine-grained clock gating has been performed to reduce the dynamic power.

Initially, OPSC decoder was synthesized with 0.7V and 0.8V supply voltages at 1.2 GHz clock frequency. Synthesis results in Table \ref{table:synthesis_voltage_results} show that 73 paths have timing violations at 0.7V. These violations can be fixed by reducing clock frequency or increasing the number of pipeline stages. The former is not possible to achieve our Tb/s throughput target. The latter increases the number of DFF and therefore complexity of clock distribution network. This may cause routing problems at backend implementation stage. Since we want to keep pipeline depth as small as possible,the backend study is performed for only OPSC decoder at 0.8V.

\begin{table}[ht]
\centering
\caption{Synthesis results of OPSC decoder at 0.7V and 0.8V supply voltages}
\label{table:synthesis_voltage_results}
\begin{tabular}{cccccc}
\hline
\begin{tabular}[c]{@{}c@{}}Supply\\ Voltage\\ (V)\end{tabular} & \begin{tabular}[c]{@{}c@{}}Cell\\ Area\\ (um$^2$)\end{tabular} & \begin{tabular}[c]{@{}c@{}}\# of \\ DFF\end{tabular} & \begin{tabular}[c]{@{}c@{}}WNS\\ (ps)\end{tabular} & \begin{tabular}[c]{@{}c@{}}TNS\\ (ps)\end{tabular} & \begin{tabular}[c]{@{}c@{}}\# of \\ violating\\ paths\end{tabular} \\ \hline
0.8 & 462,386 & 397,030 & 0 & 0 & 0 \\ 
0.7 & 468,132 & 395,935 & -5.7 & -166.3 & 73 \\ \hline
\end{tabular}
\end{table}

\subsection{Floorplan} \label{sec:asic_floorplan}
Physical floorplan for OPSC decoder is shown in Fig. \ref{fig:sc_floorplan}. Input and output pins are placed at top and bottom of the chip, respectively. Rectangular shape is adopted to increase area utilization under flat placement. The total area of the chip is $(1.255)(0.63) = 0.79$ mm$^2$.

\begin{figure}[ht]
	\centering
	\includegraphics[scale=0.45]{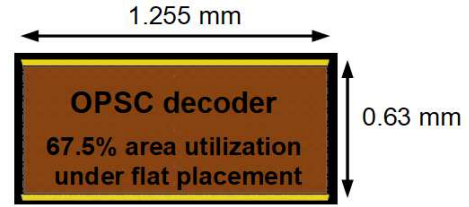}
	\caption{Physical floorplan of the OPSC decoder}
	\label{fig:sc_floorplan}
\end{figure}

\subsection{Placement and Routing} \label{sec:placement_clock_tree_routing}
Virtual silicon tape out is a technique to conceptually validate on silicon without making a real tape out. It consists of going through all the implementation and signoff phases and doing all the simulations required to validate the design without send it to the fabrication. In this study, we have completed virtual silicon design flow up to post-place-and-route stage with timing closure.

The physical implementation on this design was done with Cadence INNOVUS tool, using the exact same libraries as the synthesis and using the signoff timing  recommendations provided by TSMC for OCV. Timing is closed only on SSG0.72V and TT0.8V corners in all temperatures. We also followed all the recommendations in terms of power and layout signoff. Additional cells added through different steps of the implementation is shown in Table \ref{table:implementation_added_cells}.

\begin{table}[ht]
\centering
\caption{Additional cells at post-place, post-cts and post-route steps}
\label{table:implementation_added_cells}
\begin{tabular}{cccc} \hline
 & Post-place & Post-cts & Post-route \\ \cline{2-4}
Density & 60.48 & 66.95 & 67.49 \\
Cells to fix setup/transition & 596 & 6119 & 450 \\
Latency average & N/A & 600 ps & 600 ps \\
Cells added to fix hold & N/A & 110452 & 116018 \\ \hline
\end{tabular}
\end{table}

The spread of cells across libraries are shown in Table \ref{table:implementation_spread_cells}.

\begin{table}[ht]
\centering
\caption{Cell types of OPSC decoder}
\label{table:implementation_spread_cells}
\begin{tabular}{ccccc} \hline
\multicolumn{1}{l}{} & RVT & LVT & OPT-LVT & ULVT \\ \cline{2-5}
Number of cells & 637,149 & 207,964 & 234 & 61,618 \\
\% of   cells & 70.27 & 22.94  & 0.03  & 6.8  \\
 \hline
\end{tabular}
\end{table}

\subsection{ASIC implementation results} \label{sec:asic_results}
The implementation results of the OPSC decoder in Table \ref{table:implementation_details} show that the decoder utilizes 906K instances in 0.5 mm$^2$ cell area. Even after our optimizations, the design is still register dominated due to deeply pipelined architecture to cope with 1.2 GHz clock frequency. Since register cells are usually larger than the other cells, 70\% of the area is occupied by registers. The second largest area belongs to combinational logic to process LLRs and produce hard decisions. The total power dissipation is 1.2 W, while the leakage power is only 2.4 mW. The registers clocked at 1.2 GHz has 52.8\% of the total power dissipation. Hold-fix and setup-fix buffers also have significant low power dissipation.

\begin{table}[ht]
\centering
\caption{Implementation results of the OPSC decoder}
\label{table:implementation_details}
\begin{tabular}{ccccccc}
\cline{1-7}
 \multirow{2}{*}{Cell type} & \multicolumn{2}{c}{Instances} & \multicolumn{2}{c}{Area} & \multicolumn{2}{c}{Power} \\  
 & Value & \% & um$^2$ & \% & mW & \% \\ \hline
\multicolumn{1}{c}{Registers} & 401,132 & 44.2 & 357,327 & 69.8 & 616.2 & 52.8 \\ 
\multicolumn{1}{c}{Inverters} & 29,575 & 3.3 & 4,055 & 0.8 & 35.3 & 3.0 \\ 
\multicolumn{1}{c}{Buffers} & 154,379 & 17.0 & 52,087 & 10.2 & 261.4 & 22.4 \\ 
\multicolumn{1}{c}{Clk. latches} & 353 & 0.04 & 453 & 0.1 & 6.0 & 0.5 \\ 
\multicolumn{1}{c}{Comb. logic} & 321,292 & 35.4 & 97,683 & 19.1 & 248.6 & 21.3 \\ 
\multicolumn{1}{c}{Total} & 906,731 & 100 & 511,605 & 100 & 1,167 & 100 \\ \hline
\end{tabular}
\end{table}

\subsection{ASIC implementation comparison with other high throughput polar decoders} \label{sec:results_comparison}
A comparison of the proposed OPSC decoder implementation with the state-of-the-art polar SC decoder implementations is shown in Table \ref{table:synthesis_comparison}. The results are scaled to the same 0.8V supply voltage and 16nm process technology for a fair comparison. As for common scaling factors given in \cite{Wong2010}, the area is scaled in proportion to the square of the process ratio; the power is scaled in proportion to the square of voltage ratio and linear to process ratio; energy efficiency is scaled in proportion to the square of the process ratio times the square of the voltage ratio.

The synthesis results show that OPSC decoder has a noticeable area efficiency, energy efficiency, and latency advantage compared to others. The unrolled implementation \cite{Gross2017} provides immense throughput at high clock frequency; however, it consumes too much power. The combinational decoder implementation \cite{Dizdar2017} is favorable in terms of power and power density; however, it has extremely low throughput to satisfy the high throughput requirements of certain applications.

\begin{table}
\centering
\caption{Synthesis results comparison with the high throughput polar decoders}
\label{table:synthesis_comparison}
\begin{tabular}{@{}lccc@{}}
\hline
\textbf{Implementation}                         & This work &  \cite{Gross2017}   & \cite{Dizdar2017} \\
\textbf{ASIC Technology}                        & 16nm &   28nm  & 90nm \\ 
\textbf{Block Length}                           & 1024 &   1024  & 1024 \\ 
\textbf{Code Rate}                              & 0.83 &  0.5   & Flex. \\ 
\textbf{Supply Voltage} (V)   					& 0.8  &   1.0   & 1.3 \\ \hline
\textbf{Coded Throughput} (Gb/s)                & 1229   & 1275    & 2.6 \\
\textbf{Frequency} (MHz)                        & 1200   & 1245    & 2.5\\
\textbf{Latency} ($\mu$s)                       & 0.05  & 0.29    & 0.40$^\dagger$  \\
\textbf{Latency} (Clock Cycles)                 & 60    & 365     & 1 \\ 
\textbf{Area} (mm\textsuperscript{2})           & 0.47   & 4.63     & 3.21\\
\textbf{Power} (mW)                             & 1072 & 8793   & 191 \\ \hline
\multicolumn{4}{l}{Scaled to 16nm and 0.8V using the common scaling factors in \cite{Wong2010}. }\\ \hline
\textbf{Coded Throughput} (Gb/s)                & 1229   & \textbf{2231}    & 14.4  \\
\textbf{Frequency} (MHz)                        & 1200   & \textbf{2179}    & 14 \\
\textbf{Latency} ($\mu$s)                       & \textbf{0.05}  & 0.17    & 0.07 \\
\textbf{Area} (mm\textsuperscript{2})           & 0.47   & 1.51   & \textbf{0.10}\\
\textbf{Area Eff.} (Gb/s/mm\textsuperscript{2}) & \textbf{2590} & 1477   & 142 \\
\textbf{Power} (mW)                              & 1072  & 3216  & \textbf{13} \\
\textbf{Power Density} (mW/mm\textsuperscript{2})& 2260  & 2128    & \textbf{126} \\
\textbf{Energy Eff.} (pJ/bit)             & \textbf{0.87}   & 1.44    & 0.89 \\ \hline
\multicolumn{4}{l}{$^\dagger$Not presented in the paper, calculated from the presented results} \\
\end{tabular}
\end{table}

The comparison of the post-P\&R results of this work with the results of fabricated ASICs is shown in Table \ref{table:fabrication_comparison}. The scaled results show that this work has ultra-low latency and it can achieve Tb/s throughput under a reasonable area and power budget. The area efficiency result of this work is more than 10 times greater than the others. It is remarkable for this work to achieve 1.2 W power and 0.95 pJ/bit energy efficiency at 16nm FinFet technology.

\begin{table}
\centering
\caption{Implementation results comparison with fabricated ASICs}
\label{table:fabrication_comparison}
\begin{tabular}{@{}lccc@{}}
\hline
\textbf{Implementation}                         & This work$^\dagger$ &  \cite{Giard2017}   & \cite{Liu2018} \\
\textbf{ASIC Technology}                        & 16nm &   28nm  & 16nm \\ 
\textbf{Block Length}                           & 1024 &   1024  & 32768 \\ 
\textbf{Code Rate}                              & 0.83 &   0.85  & 0.85 \\ 
\textbf{Supply Voltage} (V)   					& 0.8  &   0.9   & 0.9 \\ \hline
\textbf{Coded Throughput} (Gb/s)                & 1229   & 9.23    & 4.44 \\
\textbf{Frequency} (MHz)                        & 1200   & 451    & 1000\\
\textbf{Latency} ($\mu$s)                       & 0.05  & 0.63    & -  \\
\textbf{Latency} (CCs)                          & 60    & 283     & - \\ 
\textbf{Area} (mm\textsuperscript{2})           & 0.79   & 0.35     & 0.35\\
\textbf{Power} (mW)                             & 1167 & 10.6   & - \\ \hline
\multicolumn{4}{l}{Scaled to 16nm and 0.8V using the common scaling factors in \cite{Wong2010}. }\\ \hline
\textbf{Coded Throughput} (Gb/s)                & \textbf{1229}   & 16.16    & 4.44  \\
\textbf{Frequency} (MHz)                        & \textbf{1200}   & 789    & 1000 \\
\textbf{Latency} ($\mu$s)                       & \textbf{0.05}  & 0.36    & - \\
\textbf{Area} (mm\textsuperscript{2})           & 0.79   & \textbf{0.11}   & 0.35\\
\textbf{Area Eff.} (Gb/s/mm\textsuperscript{2}) & \textbf{1554} & 143   & 12.7 \\
\textbf{Power} (mW)                              & 1167  & \textbf{5}  & - \\
\textbf{Power Density} (mW/mm\textsuperscript{2})& 1473  & \textbf{43}    & - \\
\textbf{Energy Eff.} (pJ/bit)             & 0.95   & \textbf{0.30}    & - \\ \hline
\multicolumn{4}{l}{$^\dagger$Post-P\&R results are given for this work.}\\
\end{tabular}
\end{table}
\section{Conclusion} \label{sec:conclusion}
In this paper, an optimized implementation of SC decoder based on AQ and R-RB methods is proposed for polar codes. These methods not only reduce implementation complexity, power and area but also enable Tb/s throughput on ASIC. Hardware architectures of (1024,854) OPSC decoder are developed for FPGA and ASIC. For FGPA, 200 Gb/s throughput is achieved and hardware verification of OPSC decoder is completed by measuring $1.1 \times 10^{-13}$ BER at 8 dB Eb/No. The ASIC implementation results show that OPSC decoder achieves 1.2 Tb/s coded throughput, 1554 Gb/s/mm\textsuperscript{2} area efficiency and 0.95 pJ/b energy efficiency. When OPSC decoder is compared with other fabricated ASICs for polar codes, it has 16 times more throughput, 7.2 times less latency and 10 times better area efficiency than the best alternative implementation.
\section*{Acknowledgment}
This work has been supported by EPIC project funded by the European Union's Horizon 2020 research and innovation programme under grant agreement No 760150. The authors would like to thank Prof. Norbert Wehn and Mr. Claus Kestel for useful discussions and comments.

\FloatBarrier
\bibliography{references}{}
\bibliographystyle{ieeetr}
\end{NoHyper}
\end{document}